\documentclass[12pt]{article}
\usepackage[a4paper,margin=2cm]{geometry}
\usepackage{cmap}
\usepackage{concmath}
\usepackage{amsmath,amssymb,textcomp}
\usepackage{cite,icomma,indentfirst,fancyvrb}
\usepackage{graphicx,overpic,wrapfig,contour}
\usepackage[labelsep=period,labelfont=bf]{caption}
\usepackage[utf8]{inputenc}
\usepackage[T1,T2A]{fontenc}
\usepackage[english,russian]{babel}

\newcommand{\code}[1]{``\texttt{#1}''}

\newcommand{\hm}[1]{#1\nobreak\discretionary{}{\hbox{\ensuremath{#1}}}{}}
\newcommand{\w}{\linewidth}

\newcommand{\Ci}[2]{\mathbf{I}_{#1}(#2)}

\newcommand{\Osr}[1]{\overline{#1\mathstrut}}

\fvset{fontsize=\small,frame=leftline,framesep=.4em,
       xleftmargin=1.2em,xrightmargin=0.2em,
       numbers=left,numbersep=.4em}
\contourlength{.2em}
\contournumber{32}

\usepackage[pdftex,unicode]{hyperref}
\hypersetup{unicode=true,
pdfauthor={P.V. Moskalev},
pdfsubject={Percolation modeling of hydraulic hysteresis in a porous media},
pdftitle={Moskalev P.V. Percolation modeling of hydraulic hysteresis in a porous media},
pdfkeywords={site percolation, square lattice, non-metric Minkowski distance, Moore neighborhood, mass fractal dimension, R programming language, SPSL package, SECP package}}

\begin{document}

\begin{titlepage}

\English
\begin{flushleft}\bfseries
UDC 519.676
\end{flushleft}
\begin{flushleft}\bfseries\Large
Percolation modeling of hydraulic hysteresis in a porous media
\end{flushleft}
\begin{flushleft}
\textbf{P.V. Moskalev}\\[1ex]
Voronezh State Agricultural University (moskalefff@gmail.com)
\end{flushleft}

\noindent\textbf{Abstract.} 
In this paper we consider various models of hydraulic hysteresis in invasive mercury porosimetry.
For simulating the hydraulic hysteresis is used isotropic site percolation on three-dimensional square lattices with $(1, d)$-neighborhood.
The relationship between the percolation model parameters and invasive porosimetry data is studied phenomenologically.
The implementation of the percolation model is based on libraries SPSL and SECP, released under license GNU GPL-3 using the free programming language R.
\medskip

\medskip

\Russian
\begin{flushleft}\bfseries
УДК 519.676
\end{flushleft}
\begin{flushleft}\bfseries\Large
Перколяционное моделирование гидравлического гистерезиса в пористой среде
\end{flushleft}
\begin{flushleft}
\textbf{П.В. Москалев}\\[1ex]
Воронежский государственный аграрный университет (moskalefff@gmail.com)
\end{flushleft}

\noindent\textbf{Аннотация.}
В работе рассматриваются различные модели гидравлического гистерезиса, возникающего при инвазивной ртутной порометрии.
Для моделирования гидравлического гистерезиса используется изотропная перколяция узлов на трёхмерных квадратных решётках с $(1, \pi)$-окрестностью.
Феноменологически исследуется взаимосвязь данных инвазивной порометрии с параметрами перколяционной модели.
Реализация перколяционной модели основана на библиотеках SPSL и SECP, выпущенных под лицензией GNU GPL-3 с использованием свободного языка программирования R.
\medskip

\noindent\textbf{Ключевые слова:} инвазивная ртутная порометрия, гидравлический гистерезис, перколяция узлов, квадратная решётка, неметрическое расстояние Минковского, окрестность Мура, массовая фрактальная размерность, язык программирования R, библиотека SPSL, библиотека SECP


\normalsize

\end{titlepage}

\section{Введение}

Одним из наиболее интересных явлений, возникающих при исследовании пористых структур, является гидравлический гистерезис.
Явление гистерезиса вообще появляется в динамических системах, состояние которых определяется совокупностью воздействий не только в текущий, но и в предшествующие моменты времени.
Наиболее ярко это свойство проявляется в системах, обладающих двумя предельными состояниями с переходной областью между ними.
Причиной появления гистерезиса обычно является асимптотический характер реакции системы на внешнее воздействие.
В результате при переменном внешнем воздействии на такую систему её текущее состояние начинает <<отставать>> от вызывающих его воздействий, что приводит к зависимости состояния от совокупности предшествующих воздействий и порождает неоднозначность параметров, характеризующих динамическую систему. 

\begin{wrapfigure}{R}{55mm}\vspace{-1em}
\centering\includegraphics[width=.95\w]{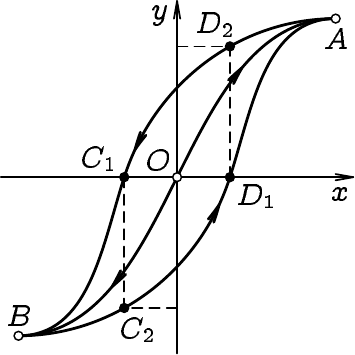}
\caption{\label{pic:hyst_loop}
Симметричные гистерезисные кривые}
\end{wrapfigure}

На рис.\,\ref{pic:hyst_loop} показан типичный пример симметричных гистерезисных кривых.
Совпадающий с началом координат центр симметрии в точке $O$ соответствует некоторому <<среднему>>, а точки $A$ и $B$~--- <<верхнему>> и <<нижнему>> предельным состояниям данной динамической системы.
Тогда в зависимости от начального состояния и направления воздействия на систему траектории её изображающих точек могут существенно различаться.
Так, при начальном состоянии системы в точке $O$ в зависимости от выбранного направления воздействия могут быть реализованы траектории $OA$ или $OB$.
Однако, при начальном состоянии системы в точке $A$ и воздействии в <<нижнем>> направлении траектория системы пройдёт по кривой $AB$, не совпадающей в промежуточных состояниях ни с траекторией $OA$, ни с $BA$.
Аналогично, при начальном состоянии системы в точке $B$ и воздействии в <<верхнем>> направлении траектория системы пройдёт по кривой $BA$, не совпадающей в промежуточных состояниях ни с траекторией $OB$, ни с $AB$.

Это приводит к тому, что знания лишь одной фазовой координаты (только $x$, или только $y$) для однозначной идентификации промежуточного состояния подобной динамической системы оказывается уже не достаточно.
Действительно, на рис.\,\ref{pic:hyst_loop} хорошо видно, что все принадлежащие отрезкам $C_1C_2$ и $D_1D_2$ точки имеют постоянные фазовые координаты $x$.
Тогда в зависимости от текущего воздействия на динамическую систему и предыстории её состояний изображающие точки $(x_1, y_1)$ или $(x_2, y_2)$ с первыми фазовыми координатами $x_1=x_C$ или $x_2=x_D$ могут иметь любые значения вторых фазовых координат из интервалов $0\leqslant y_1\leqslant y_C$ или $0\leqslant y_2\leqslant y_D$, что и порождает вышеуказанную неопределённость.

\section{Инвазивная ртутная порометрия}

Инвазивная ртутная порометрия представляет собой типичный пример комплексного физического процесса, тесно связанного с гистерезисными явлениями.
В экспериментальных исследованиях пористых структур метод инвазивной ртутной порометрии относится к числу наиболее распространённых и хорошо изученных методов.
В его основе лежит высказанная Е. Уошбурном \cite{washburn.1921.porosimetry} идея создания контролируемого перепада давления $\Delta p$ в окружающей пористое тело жидкой ртути для вдавливания некоторого её объёма $\Delta v$ в капилляры последнего.
Квазистатическое увеличение перепада давлений $\Delta p$ позволяет ртути постепенно проникать во всё более мелкие капилляры пористого тела, эквивалентный диаметр $d$ которых будет соответствовать величине вынуждающей силы $\Delta p$, а приращение удельного объёма инжектируемой жидкости $\Delta v$~--- суммарному объёму пор данного диаметра на единицу массы исследуемого образца.
Зависимость $\Delta v(\Delta p)$, получаемая в результате порометрических испытаний, называется порограммой.

Для перехода от перепада давлений $\Delta p$ к эквивалентному диаметру капилляров $d$ используется \cite{washburn.1921.porosimetry} модифицированное уравнение Юнга--Лапласа:
\begin{equation}\label{eq:washburn}
\Delta p = -\sigma \left(\frac{1}{r_1}+\frac{1}{r_2}\right) =
-\frac{4\sigma\cos\theta}{d},
\end{equation}
где $\Delta p=p_i-p_o$~--- перепад между внутренним $p_i$ и внешним $p_o$ давлениями в инжектируемой жидкости;
$\sigma$~--- коэффициент поверхностного натяжения для инжектируемой жидкости; 
например, при температурах 20--25\textcelsius\ коэффициент поверхностного натяжения $\sigma_\mathrm{Hg}$ для межфазной границы <<ртуть~--- вакуум>> составляет 0,480--0,485~Н/м;
$r_{1,2}$~--- главные радиусы кривизны мениска инжектируемой жидкости; 
$\theta$~--- краевой угол смачивания твёрдой поверхности пористого тела инжектируемой жидкостью;
$d$~--- эквивалентный диаметр капилляров пористого тела.

Очевидно, что в соотношении \eqref{eq:washburn} между перепадом давлений $\Delta p$ в инжектируемой жидкости и эквивалентным диаметром заполняемых капилляров наблюдается обратно пропорциональная зависимость $\Delta p=-\frac{4c}{d}$, где $c\hm=\sigma\cos\theta$~--- постоянная, определяемая свойствами инжектируемой жидкости.
Преимущественное использование при инвазивной порометрии жидкой ртути обусловлено тем, что краевой угол смачивания ртутью $\theta_\mathrm{Hg}$ для значительного числа природных и искусственных материалов составляет 138--156\textdegree\ со средним значением 141\textdegree\ \cite{ethington.1990.contact.angle, iupac.1994.porous}.
Помимо ртути для инвазивной порометрии может применяться и любая другая жидкость с краевым углом смачивания $\theta$ для твёрдой поверхности образца, превышающим 90\textdegree.
В частности для исследования пористых серебряных катализаторов более целесообразным представляется использование глицерина, а для песчаников и катализаторов на основе пирогенного кремнезёма~--- сплава Вуда \cite{platchenov.1988.porosimetry, borman.2000.nanopercolation}. 

Инвазивная ртутная порометрия позволяет оценивать поры с эквивалентным гидравлическим диаметром $d$ от 3,5~нм до 500~мкм \cite{giesche.2006.porosimetry}.
Для снижения влияния на результирующие данные низкой связности тупиковых пор и <<захвата>> во внутрипоровом пространстве посторонних газов и/или жидкостей пористый образец перед испытаниями подвергается вакуумированию, а вдавливаемая жидкость~--- фильтрованию и двойной перегонке для очистки от посторонних частиц и исключения газовыделения в процессе испытаний.

Одним из существенных источников ошибок метода инвазивной порометрии является необходимость обеспечения заданной точности при измерении перепадов давлений $\Delta p$ в весьма широком диапазоне.
Например, для равноточной оценки с помощью инвазивной ртутной порометрии распределения пор по размерам $d$ на интервале от 4~нм до 400~мкм в пористом теле с $\theta_\mathrm{Hg}=141$\textdegree\ при температуре 25\textcelsius\ потребуется обеспечить равноточное измерение перепадов давлений $\Delta p$ в диапазоне от 3,8 до 376915,8~кПа.

Ещё одним источником ошибок метода инвазивной жидкостной порометрии является гипотеза о постоянстве произведения коэффициента поверхностного натяжения на косинус краевого угла смачивания для всех капилляров исследуемого пористого тела $c=\sigma\cos\theta$.
Даже для макроскопически изотропных пористых тел в зависимости от условий их образования свойства поверхности в различных точках могут существенно различаться.
Это приводит к появлению ненулевой дисперсии в распределении параметра $\theta$ по внутрипоровой поверхности тела, что и порождает ошибку в определении эквивалентного диаметра капилляров $d$.

\section{Явление гидравлического гистерезиса}

Важное значение при проведении порометрических испытаний имеет характер изменения перепада давлений $\Delta p$.
Если испытания проводятся при изменении перепада от минимального $\Delta p_{\min}$ к максимальному $\Delta p_{\max}$, то полученную порограмму $\Delta v(\Delta p)$ называют кривой интрузии, а при обратном направлении изменения перепада давлений~--- кривой экструзии.
Минимальная программа порометрических испытаний включает в себя снятие обеих кривых.

\begin{wrapfigure}{R}{64mm}
\centering\includegraphics[width=.95\w]{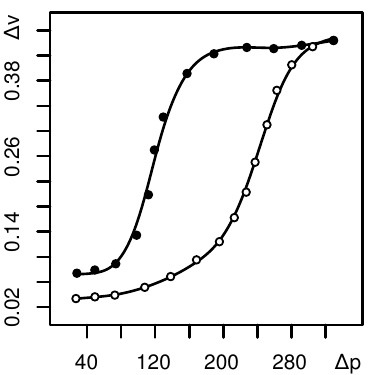}
\caption{Интегральные порограммы $\Delta v$, $\text{см}^3/\text{г}$ от $\Delta p$, МПа для силикагеля МСК-400 \cite{platchenov.1988.porosimetry}}
\label{pic:wet-dry}
\end{wrapfigure}

Примеры интегральных порограмм $\Delta v(\Delta p)$, полученные с помощью инвазивной ртутной порометрии образца мелкозернистого крупнопористого силикагеля марки МСК-400 по данным \cite{platchenov.1988.porosimetry}, приведены на рис.\,\ref{pic:wet-dry}.

Символы ``$\circ$'' и ``$\bullet$'' соответствуют кривым интрузии $\Delta v_i(\Delta p)$ и экструзии $\Delta v_e(\Delta p)$.
Значения перепадов давлений $\Delta p$ по оси абсцисс указаны в МПа, а значения удельного объёма $\Delta v$ по оси ординат~--- в $\text{см}^3/\text{г}$.
Хорошо заметно, что кривая экструзии располагается выше кривой интрузии $\Delta v_e(\Delta p)> \Delta v_i(\Delta p)$ при $\Delta p\hm< \Delta p_{\max}\hm\approx 330$~МПа, а после снижения перепада давлений до минимального уровня не возвращается в начальную точку: $\Delta v_e-\Delta v_i\hm\approx 0,04\ \text{см}^3/\text{г}$ при $\Delta p=\Delta p_{\min}\approx 27$~МПа.
Последнее свидетельствует о захвате некоторого объёма ртути во внутрипоровом пространстве образца силикателя.

Указанные несоответствия как раз и являются проявлениями гидравлического гистерезиса в пористой среде.
В современных обзорах по инвазивной ртутной порометрии выделяют три основных модели формирования гидравлического гистерезиса \cite{giesche.2002.porosimetry, giesche.2006.porosimetry}: контактного, структурного и перколяционного гистерезиса.

\subsection{Модель контактного гистерезиса}

В модели контактного гистерезиса в качестве основной причины расхождения кривых интрузии и экструзии постулируется изменение краевого угла смачивания $\theta$ при возрастании перепада давлений $\Delta p$ в инжектируемой ртути и при его убывании $\theta_i\neq \theta_e$.

\begin{figure}[tbh]
\centering\small
\begin{overpic}[width=.47\w]{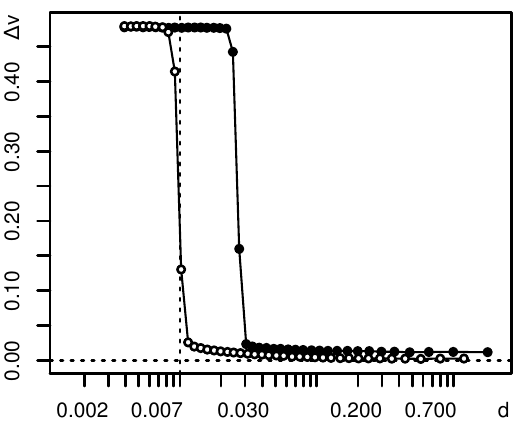}
\put(88,70){а)}
\end{overpic}\qquad
\begin{overpic}[width=.47\w]{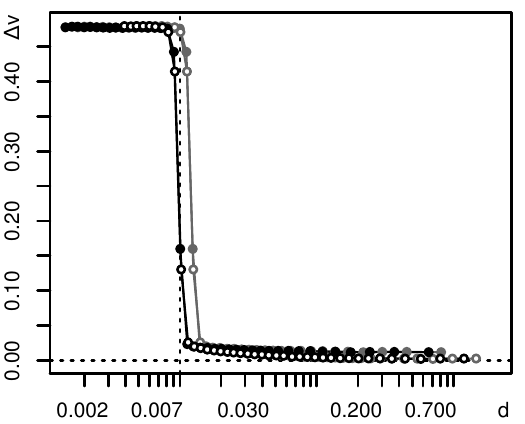}
\put(88,70){б)}
\end{overpic}
\caption{\label{pic:theta_hyst}
Результаты обработки интегральных порограмм для пористого кремнезёма по данным \cite{leon.1998.porosimetry} при: а)~$\theta_i=\theta_e=140$\textdegree; б)~$\theta_i=140$\textdegree, $\theta_e=106,5$\textdegree\ (чёрный цвет) и $\theta_i=160$\textdegree, $\theta_e=110,5$\textdegree\ (серый цвет) }
\end{figure}

На рис.\,\ref{pic:theta_hyst} (а-б) показаны примеры обработки результатов порометрических испытаний $\Delta v_i(\Delta p)$ и $\Delta v_e(\Delta p)$ для пористого кремнезёма по данным \cite{leon.1998.porosimetry} с использованием уравнения \eqref{eq:washburn} при идентичных или различных углах смачивания для кривых интрузии~--- экструзии: а)~$\theta_i=\theta_e\hm=140$\textdegree; б)~$\theta_i=140$\textdegree, $\theta_e=140-33,5=106,5$\textdegree\ (графики чёрного цвета) и $\theta_i=140+20=160$\textdegree, $\theta_e=140-29,5=110,5$\textdegree\ (графики серого цвета).
Символы ``$\circ$'' и ``$\bullet$'' соответствуют кривым интрузии $\Delta v_i(d)$ и экструзии $\Delta v_e(d)$.
Значения эквивалентных диаметров капилляров $d$ по оси абсцисс указаны в мкм, а значения удельного объёма $\Delta v$ по оси ординат~--- в $\text{см}^3/\text{г}$.

На рис.\,\ref{pic:theta_hyst} (б) хорошо видно, что использование различных углов смачивания $\theta_i\neq \theta_e$ позволяет без усложнения базового уравнения \eqref{eq:washburn} получить вполне удовлетворительное совпадение кривых интрузии~--- экструзии, но некоторые вопросы остаются открытыми.

Во-первых, совмещение кривых $\Delta v_i(d)$ и $\Delta v_e(d)$, показанных на рис.\,\ref{pic:theta_hyst} (б) чёрным цветом, достигнуто за счёт изменения значения угла смачивания $\theta_e=140-33,5=106,5$\textdegree\ при сохранении неизменным угла $\theta_i=140$\textdegree.
Того же эффекта можно достичь за счёт изменения значений обоих углов $\theta_i$ и $\theta_e$. 
К примеру, серым цветом на рис.\,\ref{pic:theta_hyst} (б) показаны кривые $\Delta v_i(d)$ и $\Delta v_e(d)$ при $\theta_i=140+20=160$\textdegree\ и $\theta_e=140-29,5=110,5$\textdegree. 
Очевидно, что в каждом случае мы будем получать различные распределения удельных объёмов пор по размерам $\Delta v(d)$, что порождает неоднозначность при интерпретации непосредственных результатов порометрических
 испытаний.

Во-вторых, применение уравнения Юнга--Лапласа с различными значениями углов смачивания $\theta_i\neq \theta_e$ с геометрической точки зрения эквивалентно простому смещению кривых $\Delta v_i(d_i)$ и/или $\Delta v_e(d_e)$ по оси абсцисс на некоторую величину $d_e-d_i$, что оставляет открытым вопрос о причинах приращения удельного объёма ртути в процессе интрузии~--- экструзии $\Delta v_e-\Delta v_i=0,0095\ \text{см}^3/\text{г}$ при $0,59\leqslant\Delta p\leqslant 0,76$~МПа.

\subsection{Модель структурного гистерезиса}

В качестве основной причины появления захваченного объёма жидкости в моделях структурного гистерезиса постулируется различие в перепадах давлений, требуемых для интрузии и экструзии несмачивающей жидкости в каналах переменного диаметра.
Поровое пространство в моделях структурного гистерезиса обычно представляется в виде совокупности изолированных пор большего эквивалентного диаметра, соединённых между собой системой каналов меньшего диаметра.
А поскольку экструзия несмачивающей жидкости в такой структуре происходит при перепадах давлений, определяемых меньшими эквивалентными диаметрами соединительных каналов, то жидкость в порах большего диаметра может задерживаться при нарушении условия неразрывности.
Это нарушение тем более вероятно, чем меньше отношение эквивалентного диаметра соединительного канала к эквивалентному диаметру поры.
Из сделанных в модели определений следует, что отношение эквивалентных диаметров соединительных каналов $d_t$ к порам $d_p$ является безразмерной величиной, ограниченной отрезком $\frac{d_t}{d_p}\in [0, 1]$ и допускающей вероятностную интерпретацию.

\begin{figure}[tbh]
\centering\small
\begin{overpic}[width=.85\w]{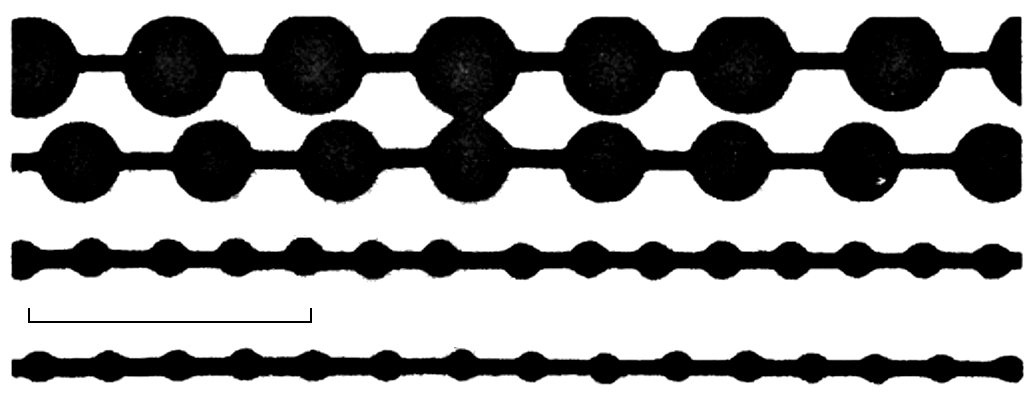}
\put(33,7.5){10 мм}
\put(93,7.5){а)}
\end{overpic}\\[4ex]
~\begin{overpic}[width=.85\w]{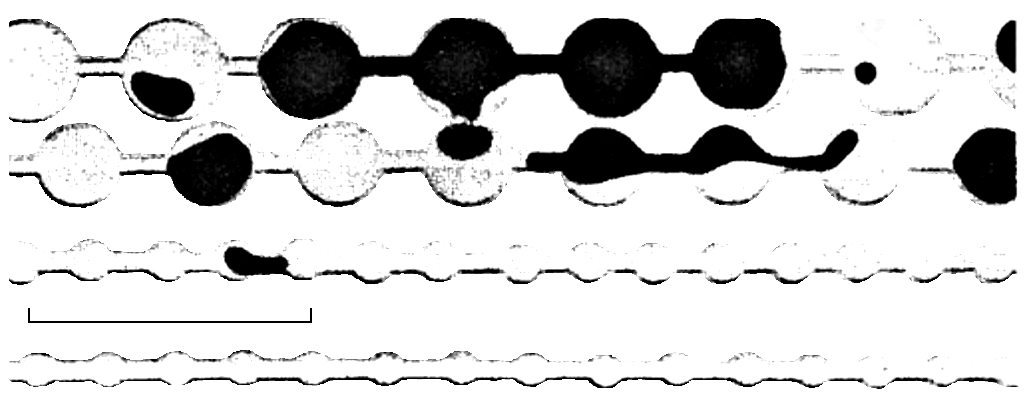}
\put(33,7.5){10 мм}
\put(93,7.5){б)}
\end{overpic}\vspace{1ex}
\caption{\label{pic:ink-bottle_hyst}
Физическая модель пористой среды с каналами переменного сечения после интрузии (а) и экструзии (б) ртути по данным \cite{wardlaw.1981.mercury} }
\end{figure}

Эффект захвата несмачивающей жидкости (ртути) был экспериментально продемонстрирован на физической модели пористой среды в работе \cite{wardlaw.1981.mercury}.
На рис.\,\ref{pic:ink-bottle_hyst} представлены проекции порового пространства этой модели, выполненной в виде нескольких последовательностей цилиндрических пор, соединённых каналами прямоугольного сечения.
Размеры масштабного отрезка соответствуют длине 10~мм. 
Нетрудно видеть, что поперечные и продольные размеры соединительных каналов постоянны для всей модели, а диаметры пор~--- в пределах каждой последовательности. 
В таком случае поперечные размеры соединительных каналов модели можно оценить как $h_t\approx 0,5$~мм, а диаметры цилиндрических пор~--- от $h_p\approx 1$~мм до $h_p\approx 3,5$~мм.
В результате оценка соотношения линейных размеров проекции наиболее узкой части канала к наиболее широкой изменяется в пределах от $\frac{h_t}{h_p}\approx 0,14$ (верхний канал на рис.\,\ref{pic:ink-bottle_hyst}) до $\frac{h_t}{h_p}\approx 0,5$ (нижний канал на рис.\,\ref{pic:ink-bottle_hyst}).

На рис.\,\ref{pic:ink-bottle_hyst} (а) показана интрузия ртути в поровое пространство физической модели при максимальном перепаде давлений, а на рис.\,\ref{pic:ink-bottle_hyst} (б)~--- экструзия ртути из порового пространства при $\Delta p=0$.
На рис.\,\ref{pic:ink-bottle_hyst} (б) хорошо видно, что даже в столь простой модели часть порового пространства после экструзии остаётся заполненной ртутью, причём эта часть оказывается тем большей, чем больше отношение эквивалентных диаметров в наиболее широкой и наиболее узкой частях данного канала.
Так, для нижних на рис.\,\ref{pic:ink-bottle_hyst} (а-б) каналов с отношением $\frac{h_t}{h_p}\approx 0,5$ видимая объёмная доля оставшейся после экструзии ртути равна нулю $\frac{v_\mathrm{Hg}}{v_0}(\Delta p=0)\approx 0$, в то время как для верхних каналов при $\frac{h_t}{h_p}\approx 0,14$ объёмная доля остаточной ртути явно превышает половину $\frac{v_\mathrm{Hg}}{v_0}(\Delta p=0)>\frac{1}{2}$.

Приведённый пример наглядно показывает, что структурный гистерезис может оказывать значимое влияние на физические процессы при инвазивной порометрии пористой среды.
В то же время используемые для приложений сетевые модели склонны к переоценке остаточного объёма несмачивающей жидкости, захватываемой в процессе испытаний пористой средой \cite{giesche.2006.porosimetry}.
Причина указанного несоответствия кроется в том, что эффект захвата несмачивающей жидкости сильно зависит от связности пористой структуры.
В пористых структурах, подобных показанной на рис.\,\ref{pic:ink-bottle_hyst}, связность невысока и для захвата ненулевого объёма несмачивающей жидкости достаточно всего лишь двух нарушений условия неразрывности.
Однако, для значительного числа испытываемых пористых материалов связность порового пространства оказывается выше модельной, что приводит к снижению объёмной доли захватываемой жидкости.

\section{Модель перколяционного гистерезиса}
\label{sec:perc_hyst}

Как было отмечено в нашей работе \cite{moskaleff.2007.porous}, топологическая структура порового пространства в случайно-неоднородной пористой среде может характеризоваться как множеством самопересечений, так и наличием частично или полностью изолированных областей, что приводит к выделению связности порового пространства в качестве одной из ключевых характеристик пористой среды.
В таком случае, для успешного моделирования физических процессов, протекающих при инвазивной порометрии, адекватное описание связности фрагментов, формирующих поровое пространство анализируемой среды, является существенно необходимым.

Для формирования модели гистерезисных явлений, возникающих при инвазивной ртутной порометрии, воспользуемся описанной в \cite{moskaleff.2013.ssTNd, moskaleff.bnak.2013.hysteresis} моделью изотропной перколяции узлов на трёхмерной квадратной решётке с $(1, \pi)$-окрест\-ностью.
Процессы интрузии~--- экструзии ртути будем моделировать для кубического образца мелкозернистого крупнопористого силикагеля марки МСК-400, интегральные порограммы $\Delta v(\Delta p)$ которого показаны на рис.\,\ref{pic:wet-dry}.

\subsection{Перколяционная решётка и граничные условия}

Примем характерный размер трёхмерной перколяционной решётки $l$, соответствующим характерному размеру элементарного репрезентативного объёма моделируемого тела, размеры которого можно определить по отношению максимального эквивалентного диаметра $d_{\max}$ к минимальному $d_{\min}$: 
$l=\frac{k_l d_{\max}}{d_{\min}}$, 
где $k_l=8\ldots 12$~--- безразмерный эмпирический коэффициент усреднения.
Используя соотношение \eqref{eq:washburn} нетрудно перейти от отношения эквивалентных диаметров $d$ к отношению соответствующих им перепадов давлений: 
$l=\frac{k_l \Delta p_{\max}}{\Delta p_{\min}}$, 
где $\Delta p_{\min}$ и $\Delta p_{\max}$~--- наименьший и наибольший перепады давлений ртути по данным порометрических испытаний.
Тогда для моделирования процессов интрузии~--- экструзии ртути в кубическом образце силикагеля МСК-400 при перепадах давлений от $\Delta p_{\min}\approx 26$~МПа до $\Delta p_{\max}\approx 330$~МПа потребуется трёхмерная квадратная перколяционная решётка с $(1, \pi)$-окрестностью размером порядка $l=\frac{10\cdot 330}{26}\approx 127$~узлов.

Учитывая, что физический процесс интрузии ртути будет распространяться от внешних границ образца внутрь пористого тела, стартовое подмножество для перколяционной модели будем формировать из достижимых узлов вдоль внешних границ решётки: $(\pm\frac{l}{2}, y, z)$; $(x, \pm\frac{l}{2}, z)$; $(x, y, \pm\frac{l}{2})$, где изменение координат $x$, $y$, $z$ ограничено интервалом $(-\frac{l}{2}, \frac{l}{2})$.

\begin{figure}[tbh]
\centering\small
\begin{overpic}[width=.47\w]{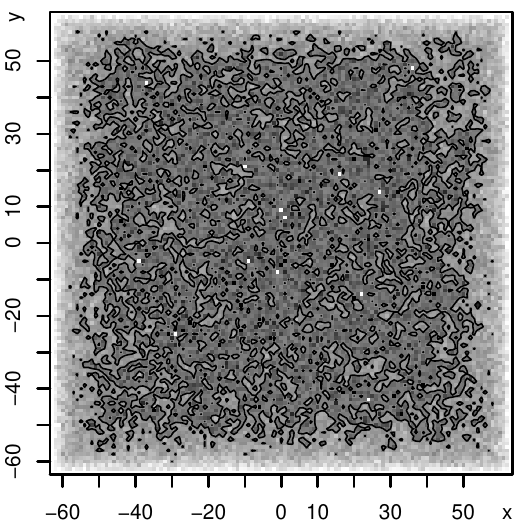}
\put(90,90){\contour{white}{a)}}
\end{overpic}\qquad
~\begin{overpic}[width=.47\w]{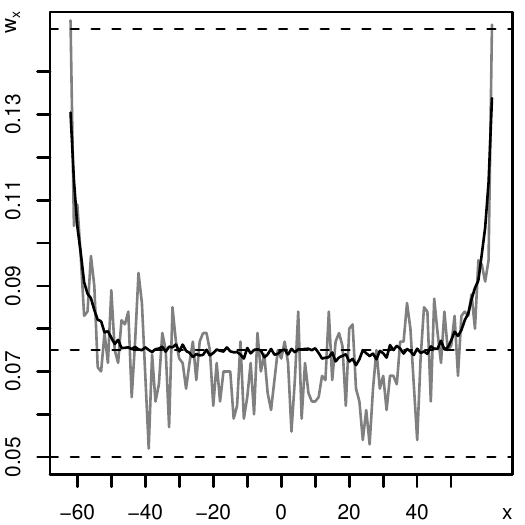}
\put(90,90){\contour{white}{б)}}
\end{overpic}\vspace{1ex}
\caption{\label{pic:perc_hyst_lattice}
Распределения относительных частот $w_{xy}$ (а) и усреднённые по оси $Oy$ распределения относительных частот $w_x$ (б) в сечении $z=0$ трёхмерной перколяционной решётки с $(1, 1)$-окрестностью при $p=0,1765$ и $n=1000$}
\end{figure}

На рис.\,\ref{pic:perc_hyst_lattice} (а) представлено распределение относительных частот узлов $w_{xy}$ в среднем горизонтальном сечении $z=0$ трёхмерной перколяционной решётки размером $l=127+2=129$ узлов с $(1, \pi)$-окрестностью по выборке объёмом $n=1000$ изотропных реализаций при показателе Минковского $\pi=1$ и доле достижимых узлов $p=0,1765$.
Увеличение ранее определённого размера решётки на два узла вызвано использованием непроницаемых граничных условий, исключающих возможность выхода при построении кластера за границу решётки, что снижает трудоёмкость статистического моделирования.
Линии уровня на рис.\,\ref{pic:perc_hyst_lattice} (а) соответствуют медианному сечению $w_{xy}=0,075$, отображённому на рис.\,\ref{pic:perc_hyst_lattice} (б) горизонтальной штриховой линией $w_x=0,075$.
Белый цвет на рис.\,\ref{pic:perc_hyst_lattice} (а) соответствует узлам с наибольшими относительными частотами $w_{xy}=0,15$, а чёрный~--- узлам с наименьшими относительными частотами $w_{xy}=0,05$.
Все узлы с относительными частотами, выходящими за пределы указанного интервала, на рис.\,\ref{pic:perc_hyst_lattice} (а) условно не показаны.
На рис.\,\ref{pic:perc_hyst_lattice} (б) верхняя и нижняя границы интервала отображаемых частот показаны горизонтальными штриховыми линиями $w_x=0,05$ и $w_x=0,15$.

Сплошными линиями на рис.\,\ref{pic:perc_hyst_lattice} (б) показаны  усреднённые по оси $Oy$ относительные частоты узлов $w_x$ в среднем сечении $z=0$ (ломаная чёрного цвета), а также распределение относительных частот узлов $w_x$ в сечении $z=0$ и $y=0$ (ломаная серого цвета).
Хорошо заметно, что медианная оценка уровня $w_x$ оказывается устойчивой к локальным максимумам относительных частот во внутренней окрестности внешней границы решётки.
На рисунках не показаны распределения относительных частот узлов $w_{yz}$ и $w_{xz}$ в средних вертикальных сечениях $x=0$ и $y=0$ перколяционной решётки, но, с учётом пространственной изотропности выборочной совокупности кластеров, значимых различий в их статистических характеристиках не ожидается.

В приведённых выше расчётах были использованы функция \code{ssi3d()} из состава библиотеки SPSL \cite{moskalev.cran.2012.spsl}, выпущенной автором под лицензией GNU GPL-3 и доступной для свободной загрузки через систему репозиториев CRAN.

\subsection{Основные гипотезы и допущения}
\label{sec:hyst_hypotesis}

При минимальных перепадах давлений $\Delta p\approx\Delta p_{\min}$ могут быть заполнены лишь наиболее крупные капилляры, эквивалентный диаметр которых будет сопоставим с размерами перколяционной решётки, а соответствующая им доля достижимых узлов~--- близка к нулю $p\approx 0$. 
По мере увеличения $\Delta p$ эквивалентный диаметр заполняемых капилляров падает, а соответствующая им доля достижимых узлов~--- возрастает.
При максимальных перепадах давлений $\Delta p\approx\Delta p_{\max}$ заполняемыми становятся даже наиболее мелкие капилляры, эквивалентный диаметр которых сопоставим с радиусом единичной окрестности перколяционной решётки, а соответствующая им доля достижимых узлов~--- близка к единице $p\approx 1$.
Тогда, с учётом равномерности взвешивающего распределения узлов перколяционной решётки $u_{xyz}\sim \mathbf{U}(0, 1)$ можно предположить, что доля достижимых узлов $p$ будет  соответствовать нормированному перепаду давлений ртути $\Delta p$ при проведении порометрических испытаний $p\sim \frac{\Delta p-\Delta p_{\min}}{\Delta p_{\max}-\Delta p_{\min}}$.
Так же как и перепад давлений $\Delta p$ доля достижимых узлов $p$ является неаддитивной величиной, которая с точностью до статистической ошибки сохраняет своё значение при разбиении исходной системы (перколяционной решётки) на части.

Интегральная кривая интрузии $\Delta v_i(\Delta p)$ показывает, как возрастает удельный объём связанного с внешним периметром порового пространства $\Delta v_i$ образца при изменении перепада давлений в инжектируемой ртути от минимального до максимального $\Delta p_{\min}\hm\leqslant \Delta p\leqslant \Delta p_{\max}$.
Аналогично, интегральная кривая экструзии $\Delta v_e(\Delta p)$ показывает, как убывает удельный объём связанного с внешним периметром порового пространства $\Delta v_e$ образца при изменении перепада давлений в ртути от максимального до минимального $\Delta p_{\max}\geqslant \Delta p\geqslant \Delta p_{\min}$.

Эмпирическая нормировка приращений связанного с внешней поверхностью объёма порового пространства $\Delta V_c$ по массе образца $M_t$ по всей видимости была выбрана из соображений простоты и точности выполнения измерений $\Delta v=\frac{\Delta V_c}{M_t}$.
При макроскопически изотропной структуре анализируемого материала его кажущаяся плотность является постоянной величиной $\rho_t=\mathrm{const}$ и масса образца $M_t$ будет прямо пропорциональна его объёму $M_t=\rho_t V_t$.
В моделях решёточной перколяции приращение эффективного объёма порового пространства $\Delta V_c$ оказывается пропорциональным приращению числа узлов кластера $\Delta N_c$, образуемого из подмножества достижимых узлов решётки, связанных с заданным стартовым подмножеством $\Delta V_c=V_{\min} \Delta N_c$, где $V_{\min}=\frac{V_t}{N_t}$~--- объём образца пористого тела, соответствующий единичному узлу решётки; $N_t$~--- общее число узлов перколяционной решётки.
Тогда, массовая фрактальная размерность $d$, определяемая приращением числа узлов изотропного кластера $\Delta N_c$, усреднённого по приращениям характерных размеров $\Delta r$ элементов покрывающего перколяционную решётку множества, будет соответствовать нормированному приращению удельного эффективного объёма порового пространства $d\sim \frac{\Delta v- \Delta v_{\min}}{\Delta v_{\max}- \Delta v_{\min}}$.

\subsection{Построение модели для отдельных реализаций}

Для статистически изотропных реализаций кластеров узлов на трёхмерной квадратной решётке неподвижная точка покрывающего множества должна быть расположена в центре решётки $(0, 0, 0)$. 
Для оценки зависимости массовой фрактальной размерности $d$ от доли достижимых узлов $p$ воспользуемся вначале отдельными реализациями изотропных кластеров узлов с $(1, \pi)$-окрестностью при $\pi=1$ и стартовым подмножеством вдоль внешних границ трёхмерной квадратной решётки размером $l=129$~узлов. 

\begin{figure}[hbt]
\centering\small
\begin{overpic}[width=.47\w]{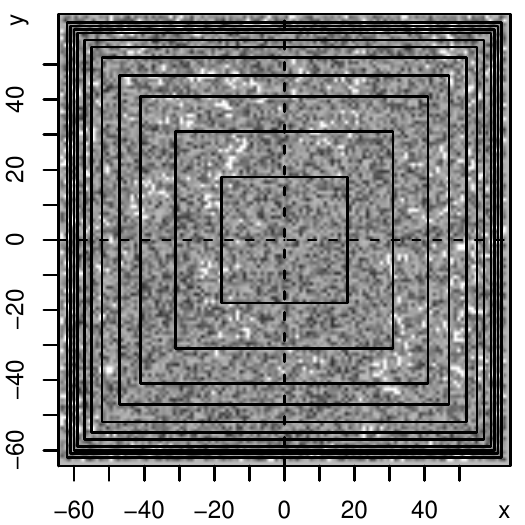}
\put(88,88){\contour{white}{a)}}
\end{overpic}\qquad
\begin{overpic}[width=.47\w]{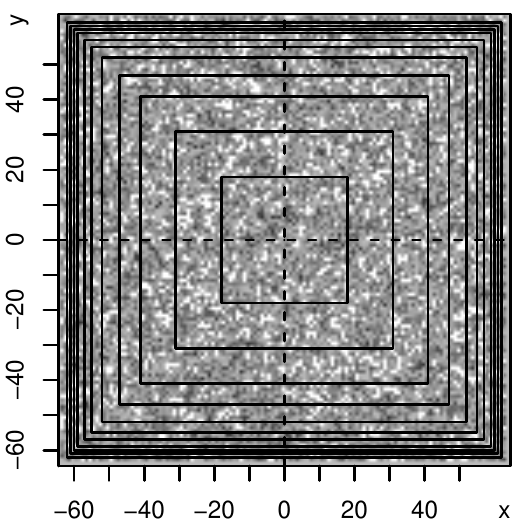}
\put(88,88){\contour{white}{б)}}
\end{overpic}
\caption{\label{pic:ifdc3s}
Покрытие реализаций изотропных кластеров узлов с $(1, 1)$-окрестностью на трёхмерной квадратной решётке размером $l=129$ узлов при: а)~$p=0,1565<p_c(1)$; б)~$p=0,1965>p_c(1)$}
\end{figure}

На рис.\,\ref{pic:ifdc3s} (а-б) показаны средние сечения $z=0$ трёхмерных реализаций изотропных кластеров для до- и сверхкритических значений долей достижимых узлов: а)~$p=0,1565<p_c(\pi=1)$; б)~$p=0,1965>p_c(\pi=1)$. 
Тогда, при стартовом подмножестве, состоящем из достижимых узлов вдоль внешних границ решётки: $(\pm\frac{l}{2}, y, z)$; $(x, \pm\frac{l}{2}, z)$; $(x, y, \pm\frac{l}{2})$, покрывающее множество будет строиться из оболочек кубических элементов с фиксированным внешним размером $l_0$ и переменной толщиной $r_i=l_0-l_i$, где $l_i$~--- переменный размер исключаемой области.
Практически это означает, что выборка абсолютных частот $\{n_i\}$ покрываемых узлов кластера будет определяться разностью абсолютных частот всех узлов перколяционной решётки и узлов, покрываемых масштабируемой областью исключения.
В примерах, показанных на рис.\,\ref{pic:ifdc3s}, множество областей исключения образуется двенадцатью симметрично сжимаемыми кубическими элементами с неподвижной точкой в центре решётки $(0, 0, 0)$.

\begin{figure}[hbt]
\centering\small
\begin{overpic}[width=.47\w]{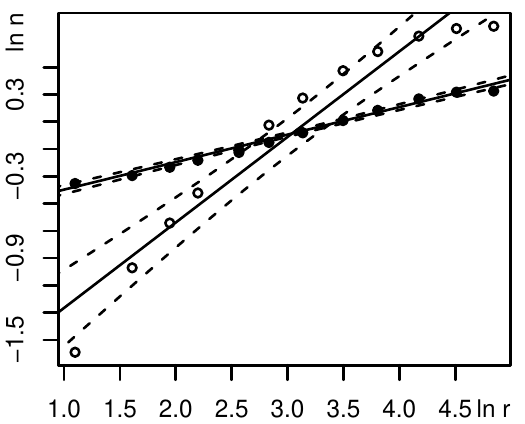}
\put(88,16){а)}
\end{overpic}\qquad
\begin{overpic}[width=.47\w]{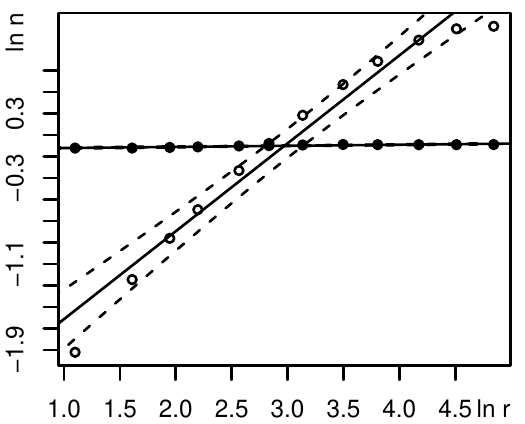}
\put(88,16){б)}
\end{overpic}\\[3ex]
\begin{overpic}[width=.47\w]{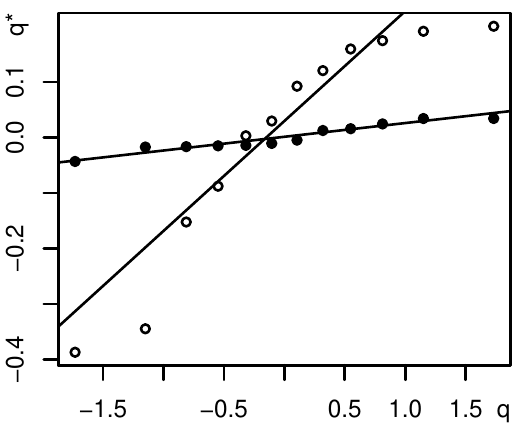}
\put(88,16){в)}
\end{overpic}\qquad
\begin{overpic}[width=.47\w]{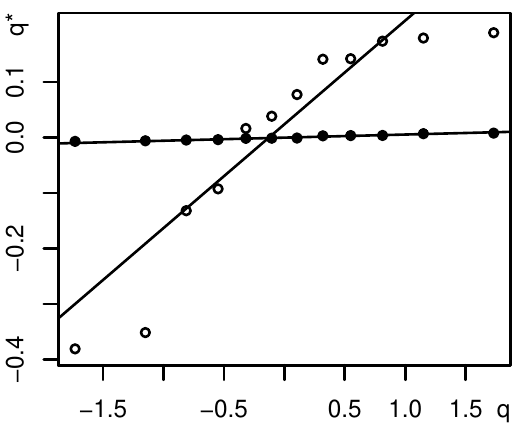}
\put(88,16){г)}
\end{overpic}
\caption{\label{pic:ifdc3s_reg}
Линейные регрессионные модели (а-б) и квантильные графики остатков (в-г), используемые при оценке фрактальной размерности для показанных на рис.\,\ref{pic:ifdc3s} реализаций изотропных кластеров}
\end{figure}

В листинге~1 показана реализация на языке R функции \code{isc3s()} из состава библиотеки <<SECP>> \cite{moskalev.cran.2012.secp}, обеспечивающей построение изотропного покрывающего множества требуемого объёма \code{k} для трёхмерной квадратной решётки заданного размера \code{x}, с неподвижной точкой \code{o}.

\paragraph{Листинг 1.} Реализация функции \code{isc3s()} на языке R
\begin{Verbatim} 
isc3s <- function(k=12, x=rep(95, times=3), o=(x+1)/2, r=min(o-2)^(seq(k)/k)) {
  r <- unique(round(r))
  return(rbind(r=2*r+1,
               x1=o[1]-r, x2=o[1]+r, 
               y1=o[2]-r, y2=o[2]+r, 
               z1=o[3]-r, z2=o[3]+r))
}
\end{Verbatim}

На рис.\,\ref{pic:ifdc3s_reg} (а-б) показаны линейные модели, позволяющие получить оценки коэффициентов линейной регрессии: а)~$\Ci{0,95}{\rho_n}=(0,506; 0,748)$ при $p=0,1565< p_c(1)$; б)~$\Ci{0,95}{\rho_n}=(0,700; 0,937)$ при $p=0,1965> p_c(1)$.
Символы ``$\circ$'' соответствуют выборочным значениям $(\ln r_i, \ln n_i)$ для суммарных абсолютных частот узлов $n_i$, покрываемых оболочкой текущего размера $r_i$.

В листинге~2 показана реализация на языке R функции \code{fdc3s()} из состава библиотеки <<SECP>> \cite{moskalev.cran.2012.secp}, обеспечивающей статистическую оценку массовой фрактальной размерности заданной реализации кластера узлов на трёхмерной квадратной решётке \code{acc} с заданным покрывающим множеством \code{bnd}.

\paragraph{Листинг 2.} Реализация функции \code{fdc3s()} на языке R
\begin{Verbatim} 
require(SPSL)
fdc3s <- function(acc=ssi30(x=95), bnd=isc3s(k=12, x=dim(acc))) {
  n <- rep(0, times=ncol(bnd))
  for (i in seq(ncol(bnd)))
    n[i] <- sum(acc[bnd["x1",i]:bnd["x2",i],
                    bnd["y1",i]:bnd["y2",i],
                    bnd["z1",i]:bnd["z2",i]] > 1)
  r <- log(bnd["r",])
  n <- log(n)
  return(lm(n ~ r))
}
\end{Verbatim}

Полученные в \cite{moskaleff.2011.fds.ssi20} результаты, позволяют предположить, что при увеличении доли достижимых узлов $p\in(0, 1)$ точечные оценки коэффициента регрессии абсолютных частот будут монотонно возрастать на промежутке $\rho_n\in(0, 1)$.
При этом толщина оболочки $r_i$ будет возрастать по $i$ как функция, выпуклая вниз, а численность узлов $n_i$~--- как функция, выпуклая вверх, что обусловлено наличием ограничения на текущее число узлов $n_i\leqslant l_0^3$ по мере увеличения толщины оболочки $r_i\leqslant l_0$ при фиксированном размере решётки $l_0$.
В результате, несмотря на достаточно высокие значения коэффициентов детерминации $R_a^2> 0,8$, адекватность линейной модели при $p\to 1-$ будет оставаться слабой.

Для решения указанной проблемы вместо суммы абсолютных частот узлов оболочки $n_i$ воспользуемся суммой их отклонений $\Delta n_i=n_x-n_i$ от предельного числа узлов $n_x$ для оболочки текущего размера $r_i$ при $p\to 1-$.
Линейные модели, позволяющие получить оценки коэффициентов регрессии отклонений, показаны на рис.\,\ref{pic:ifdc3s_reg}: а)~$\Ci{0,95}{\rho_{\Delta n}}=(0,187; 0,216)$ при $p=0,1565< p_c(\pi=1)$; б)~$\Ci{0,95}{\rho_{\Delta n}}=(0,007; 0,013)$ при $p=0,1965\hm> p_c(\pi=1)$.
Символы ``$\bullet$'' соответствуют выборочным значениям $(\ln r_i$; $\ln \Delta n_i)$ для отклонений суммарных абсолютных частот узлов $\Delta n_i$, покрываемых оболочкой текущего размера $r_i$.


По рис.\,\ref{pic:ifdc3s_reg} (а-б) можно предположить, что адекватность линейной модели выборочным данным $(\ln r_i, \ln \Delta n_i)$ будет возрастать при $p\to 1-$, а радиусы доверительных интервалов для коэффициента регрессии $\rho_{\Delta n}$~--- снижаться до $\varepsilon_{\Delta n}\to 0+$.
Действительно, оценивая нормальность распределения остатков $e_{n_i}$ и $e_{\Delta n_i}$ линейных моделей по приведённым на рис.\,\ref{pic:ifdc3s_reg} (в-г) квантиль-квантильным графикам нетрудно заметить, что показанное символами ``$\bullet$'' распределение остатков $e_{\Delta n_i}$ значительно лучше соответствует показанному сплошными линиями теоретическому нормальному распределению, чем распределение остатков $e_{n_i}$, показанное символами ``$\circ$''.
Кроме того, с учётом сделанных определений можно показать, что при увеличении доли достижимых узлов на $p\in(0, 1)$ точечные оценки коэффициента регрессии $\rho_{\Delta n}$ отклонений абсолютных частот $\Delta n$ будут монотонно убывать на промежутке $\rho_{\Delta n}\in(0, 1)$.

\subsection{Построение модели для выборочных совокупностей}

Статистические оценки коэффициента регрессии $\rho_{\Delta n}$, полученные для отдельных реализаций при докритических долях достижимых узлов решётки $p<p_c(\pi)$, будут зависеть от репрезентативности выбранной реализации, что существенно снижает ценность полученных результатов.
Как было показано в \cite{moskaleff.2011.fds.ssi20}, одним из эффективных способов для преодоления указанного недостатка является переход от оценки коэффициента $\rho_{\Delta n}$ по отклонениям абсолютных частот для отдельных реализаций кластеров к оценке коэффициента $\rho_{\Delta v}$ по отклонениям относительных частот для их выборочной совокупности.

\begin{figure}[hbt]
\centering\small
\begin{overpic}[width=.47\w]{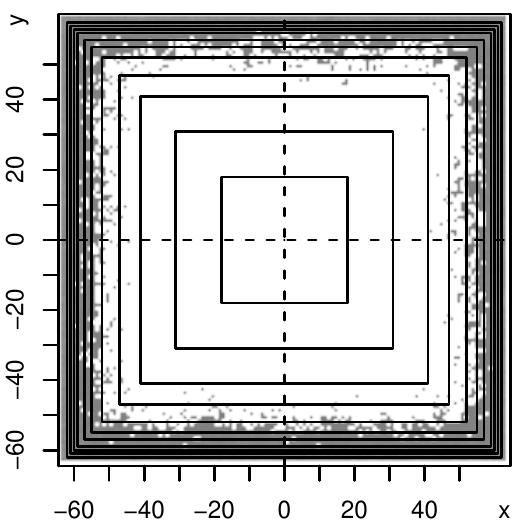}
\put(88,88){\contour{white}{a)}}
\end{overpic}\qquad
\begin{overpic}[width=.47\w]{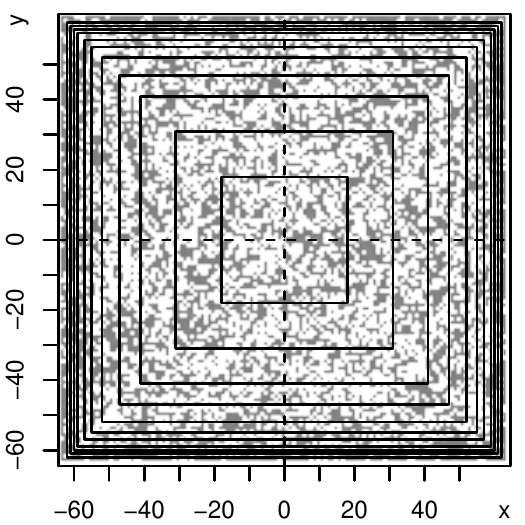}
\put(88,88){\contour{white}{б)}}
\end{overpic}
\caption{\label{pic:ifds3s}
Покрытие выборки объёмом $n=200$ реализаций изотропных кластеров узлов с $(1, 1)$-окрестностью на трёхмерной решётке размером $l=129$ узлов при: а)~$p=0,1565<p_c(1)$; б)~$p=0,1965>p_c(1)$}
\end{figure}

Тогда для оценки зависимости массовой фрактальной размерности $d$ от доли достижимых узлов $p$ воспользуемся выборочными совокупностями реализаций изотропных кластеров узлов с $(1, \pi)$-окрестностью при $\pi=1$ и стартовым подмножеством вдоль внешних границ трёхмерной квадратной решётки размером $l=129$~узлов. 
На рис.\,\ref{pic:ifds3s} (а-б) показаны средние сечения $z=0$ распределений относительных частот узлов трёхмерной перколяционной решётки по выборкам объёмом $n=200$ реализаций изотропных кластеров для до- и сверхкритических значений долей достижимых узлов: а)~$p=0,1565\hm< p_c(\pi=1)$; б)~$p=0,1965>p_c(\pi=1)$.
Белый цвет на рис.\,\ref{pic:ifds3s} (а-б) соответствует наибольшему значению относительной частоты текущего узла решётки $v=1$, а тёмно-серый цвет~--- усреднённому по выборке значению относительной частоты $\Osr{v}(p)$, которая в данном случае будет функцией доли достижимых узлов решётки: а)~$\Osr{v}(0,1565)\approx 0,0185$; б)~$\Osr{v}(0,1965)\approx 0,1265$.


В листинге~3 показана реализация на языке R функции \code{fds3s()} из состава библиотеки <<SECP>> \cite{moskalev.cran.2012.secp}, обеспечивающей статистическую оценку массовой фрактальной размерности для заданного распределения относительных частот узлов на трёхмерной квадратной решётке \code{rfq} с заданным покрывающим множеством \code{bnd}.

\paragraph{Листинг 3.} Реализация функции \code{fds3s()} на языке R
\begin{Verbatim} 
require(SPSL)
fds3s <- function(rfq=fssi30(x=95), bnd=isc3s(k=12, x=dim(rfq))) {
  w <- rep(0, times=ncol(bnd))
  for (i in seq(ncol(bnd)))
    w[i] <- sum(rfq[bnd["x1",i]:bnd["x2",i],
                    bnd["y1",i]:bnd["y2",i],
                    bnd["z1",i]:bnd["z2",i]])
  r <- log(bnd["r",])
  w <- log(w)
  return(lm(w ~ r))
}
\end{Verbatim}

\begin{figure}[hbt]
\centering\small
\begin{overpic}[width=.47\w]{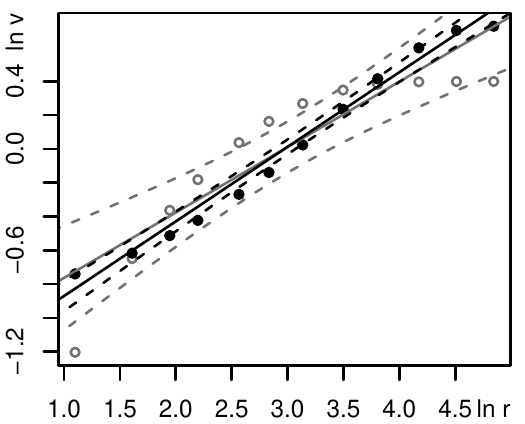}
\put(88,16){а)}
\end{overpic}\qquad
\begin{overpic}[width=.47\w]{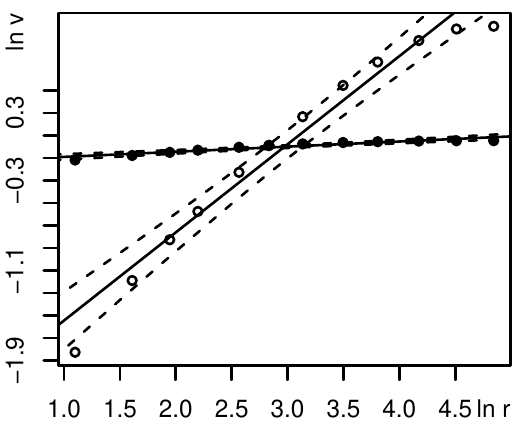}
\put(88,16){б)}
\end{overpic}\\[3ex]
\begin{overpic}[width=.47\w]{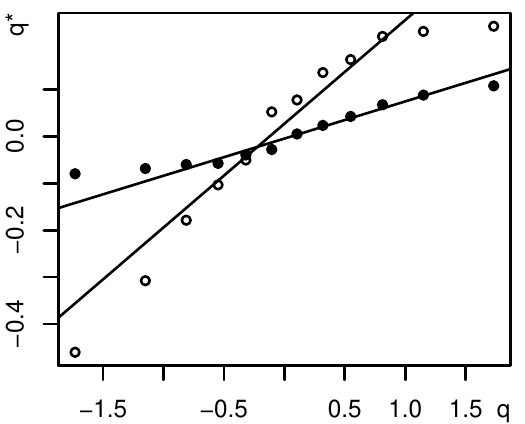}
\put(88,16){в)}
\end{overpic}\qquad
\begin{overpic}[width=.47\w]{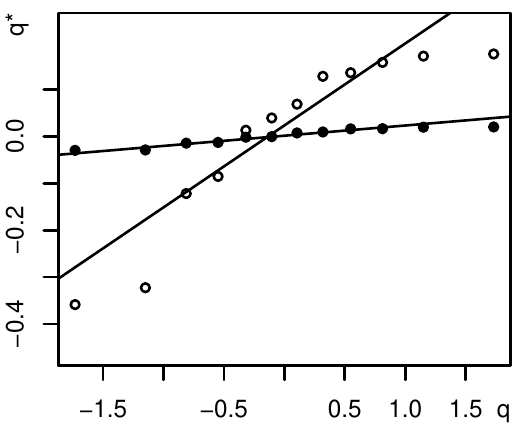}
\put(88,16){г)}
\end{overpic}
\caption{\label{pic:ifds3s_reg}
Линейные регрессионные модели (а-б) и квантильные графики остатков (в-г), используемые при оценке фрактальной размерности для показанных на рис.\,\ref{pic:ifds3s} выборок реализаций изотропных кластеров}
\end{figure}

На рис.\,\ref{pic:ifds3s_reg} (а-б) показаны линейные модели, позволяющие получить интервальные оценки коэффициентов линейной регрессии: а)~$\Ci{0,95}{\rho_v}\hm= (0,253; 0,520)$ при $p=0,1565< p_c(\pi=1)$; б)~$\Ci{0,95}{\rho_v}=(0,672; 0,893)$ при $p=0,1965> p_c(\pi=1)$.
Символы ``$\circ$'' соответствуют выборочным значениям $(\ln r_i; \ln v_i)$ для суммарных относительных частот узлов $v_i$, покрываемых оболочной текущего размера $r_i$.


Как и ожидалось, неопределённость интервальных оценок $\Ci{0,95}{\rho_v}$, найденных для выборочных совокупностей объёмом $n=200$ реализаций, по сравнению с отдельными реализациями изотропных кластеров узлов значимо не изменилась.
Действительно, $\varepsilon_n(0,1565)\approx 0,1242$, что вполне сопоставимо с $\varepsilon_v(0,1565)\approx 0,1334$, а $\varepsilon_n(0,1965)\approx 0,1184$, что также сопоставимо с $\varepsilon_v(0,1965)\approx 0,1104$.
Это подтверждает гипотезу о недостаточной адекватности линейной функции при аппроксимации эмпирических точек $(\ln n_i; \ln r_i)$ или $(\ln v_i; \ln r_i)$.

Для решения указанной проблемы вместо суммы относительных частот узлов оболочки $v_i$ воспользуемся суммой их отклонений $\Delta v_i=v_x-v_i$ от предельной суммы частот узлов $v_x$ для оболочки текущего размера $r_i$ при $p\to 1-$.
На рис.\,\ref{pic:ifds3s_reg} (а-б) показаны линейные модели, позволяющие получить интервальные оценки коэффициентов регрессии отклонений: а)~$\Ci{0,95}{\rho_{\Delta v}}\hm=(0,403; 0,480)$ при $p=0,1565< p_c(\pi=1)$; б)~$\Ci{0,95}{\rho_{\Delta v}}\hm=(0,035; 0,057)$ при $p=0,1965> p_c(\pi=1)$.
Символами ``$\bullet$'' показаны выборочные точки $(\ln r_i; \ln \Delta v_i)$ для отклонений суммарных относительных частот узлов $v_i$, покрываемых оболочной текущего размера $r_i$.

По рис.\,\ref{pic:ifds3s_reg} (а-б) также можно предположить, что адекватность линейной модели выборочным данным $(\ln r_i; \ln \Delta v_i)$ при $p\to 1-$ будет возрастать, а радиусы доверительных интервалов для коэффициента регрессии $\rho_{\Delta v}$~--- снижаться до $\varepsilon_{\Delta v}\to 0+$.
Действительно, оценивая нормальность распределения остатков $e_{v_i}$ и $e_{\Delta v_i}$ линейных моделей по приведённым на рис.\,\ref{pic:ifds3s_reg} (в-г) квантиль-квантильным графикам нетрудно заметить, что показанное символами ``$\bullet$'' распределение остатков $e_{\Delta v_i}$ по-прежнему лучше соответствует показанному сплошными линиями теоретическому нормальному распределению, чем распределение остатков $e_{v_i}$, показанное символами ``$\circ$''.
Кроме того, с учётом сделанных определений также можно показать, что при увеличении доли достижимых узлов на $p\in(0, 1)$ точечные оценки коэффициента регрессии $\rho_{\Delta v}$ отклонений относительных частот $\Delta v$ будут монотонно убывать на промежутке $\rho_{\Delta v}\in(0, 1)$.

\subsection{Моделирование перколяционного гистерезиса}

Воспользуемся описанной в предыдущих разделах методикой для анализа зависимости интервальных оценок коэффициента регрессии отклонений относительных частот $(\rho\pm\varepsilon)_{\Delta v}(p|\pi)$ от доли достижимых узлов $p$ и показателя Минковского $\pi$ по выборочной совокупности объёмом $n=200$ реализаций изотропных кластеров узлов с $(1, \pi)$-окрестностью и стартовым подмножеством вдоль внешних границ трёхмерной квадратной решётки размером $l=129$~узлов. 
Из сделанных определений следует, что статистические оценки для центров $\rho_{\Delta v}(p|\pi)$ и радиусов $\varepsilon_{\Delta v}(p|\pi)$ доверительных интервалов коэффициента регрессии отклонений относительных частот $\Delta v$ оболочек формируемых в указанных условиях кластеров будут монотонно убывать как по $p$, так и по $\pi$ от некоторого максимального значения $\rho_{\Delta v}(p|\pi)\to \rho_{\max}>0$ при $p\to 0+$ и $\pi\to 0+$ до нуля $\rho_{\Delta v}(p|\pi)\to 0+$ при $p\to 1-$ и $\pi\to \infty$.

\begin{figure}[hbt]
\centering\small
\begin{overpic}[width=.47\w]{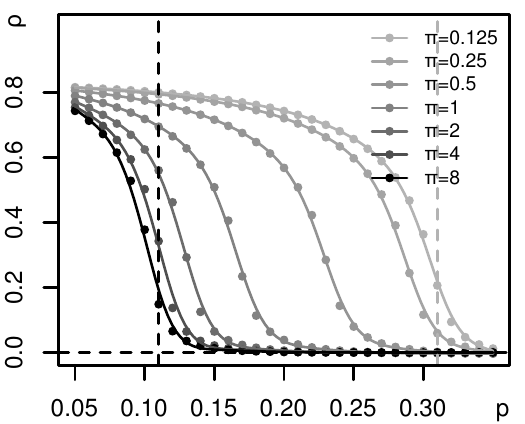}
\put(88,22){а)}
\end{overpic}\qquad
\begin{overpic}[width=.47\w]{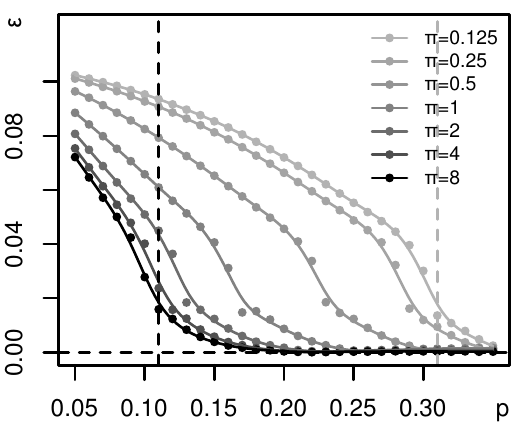}
\put(88,22){б)}
\end{overpic}
\caption{\label{pic:ssi3d_rho_ppi}
Зависимости для 0,95-доверительных интервалов коэффициента регрессии $(\rho\pm\varepsilon)_{\Delta v}$ от доли достижимых узлов $p$ и показателя Минковского $\pi$ по выборке объёмом $n=200$ реализаций кластеров узлов с $(1, \pi)$-окрестностью}
\end{figure}

На рис.\,\ref{pic:ssi3d_rho_ppi} показаны зависимости для центров (а) и радиусов (б) 0,95-доверительных интервалов коэффициента регрессии $\Ci{0,95}{\varrho}=(\rho\pm\varepsilon)_{\Delta v}$ отклонений относительных частот $\Delta v$ от доли достижимых узлов $p$ и показателя Минковского $\pi$ по выборке объёмом $n=200$ реализаций изотропных кластеров узлов с $(1, \pi)$-окрестностью при $p=0,05; 0,06;\ldots; 0,35$ и $\pi=2^{-3}, 2^{-2},\ldots, 2^3$.
Символы ``$\bullet$'' чёрного цвета на рис.\,\ref{pic:ssi3d_rho_ppi} (а-б) соответствуют зависимостям $\rho(p|\pi)$ и $\varepsilon(p|\pi)$ при наибольшем значении показателя Минковского $\rho(p|8)$, а светло-серого цвета~--- при его наименьшем значении $\rho(p|\frac{1}{8})$.
При этом вертикальная штриховая линия чёрного цвета соответствует критическому значению доли достижимых узлов $p_c(\pi)$ при наибольшем значении показателя Минковского $p_c(8)\approx 0,11$, а светло-серого цвета~--- при его наименьшем значении $p_c(\frac{1}{8})\approx 0,31$.

\begin{figure}[hbt]
\centering\small
\begin{overpic}[width=.47\w]{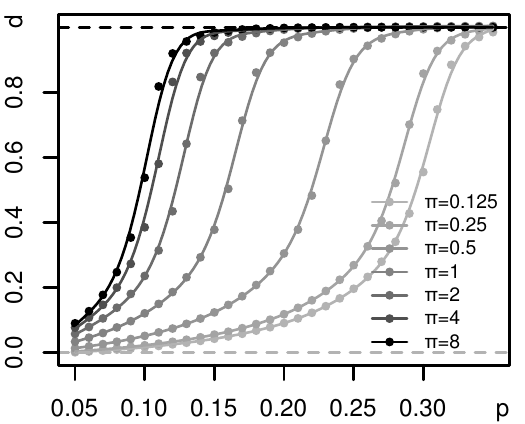}
\put(88,66){а)}
\end{overpic}\qquad
\begin{overpic}[width=.47\w]{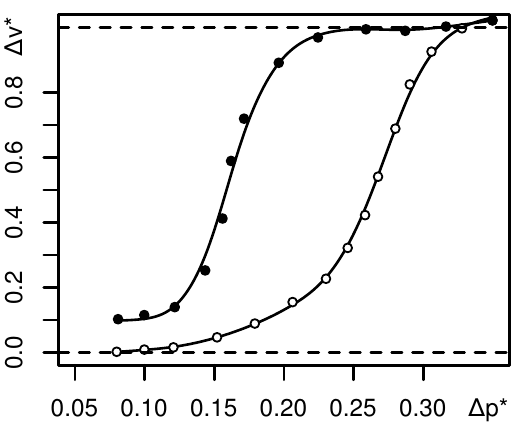}
\put(88,66){б)}
\end{overpic}
\caption{\label{pic:ssi3d_d_ppi}
Нормированные результаты перколяционного моделирования и порометрических испытаний: а)~зависимости $d_{\Delta v}(p|\pi)$; б)~зависимости $\Delta p^*(\Delta v^*)$}
\end{figure}

Для перехода от оценки коэффициента регрессии к оценке массовой фрактальной размерности заметим, что рост оболочки при масштабировании области исключения ограничен лишь одним измерением~--- её толщиной $r$. 
Как уже было показано в наших работах \cite{moskaleff.2011.fds.ssi20, moskaleff.vsuet.2012.p-sic} это приводит к получению монотонно возрастающей по $p$ оценки фрактальной размерности $d$, ограниченной интервалом $d\in [0, 1]$, с граничными значениями $d\to 0+$ при $p\to 0+$ и $d\to 1-$ при $p\to 1-$.
С учётом вышеизложенного, искомую оценку можно будет определить как нормированное отклонение статистической оценки коэффициента регрессии отклонений относительных частот $\rho_{\Delta v}$ от достигаемого при минимальном $p$ максимального значения $\rho_{\max}$: $d_{\Delta v}=\frac{\rho_{\max}-\rho_{\Delta v}}{\rho_{\max}}\in [0, 1]$. 
Нетрудно проверить, что в соответствии с ранее полученными результатами новая оценка $d_{\Delta v}$ также обеспечивает монотонный рост по $p$ с граничными значениями $d_{\Delta v}\to 0+$ при $p\to 0+$ и $d_{\Delta v}\to 1-$ при $p\to 1-$.
Полученные в результате нормировки зависимости $d_{\Delta v}(p|\pi)$ показаны на рис.\,\ref{pic:ssi3d_d_ppi} (а). 

Для сопоставления результатов моделирования с данными порометрических испытаний произведём нормировку перепадов давлений $\Delta p^*=0,08+0,27\,\frac{\Delta p-\Delta p_{\min}}{\Delta p_{\max}-\Delta p_{\min}}$ и приращений удельного объёма пор $\Delta v^*=\frac{\Delta v-\Delta v_{\min}}{\Delta v_{\max}-\Delta v_{\min}}$, как это было показано в разделе \ref{sec:hyst_hypotesis}.
Нормированные кривые $\Delta p^*(\Delta v^*)$ интрузии и экструзии ртути для образца силикагеля МСК-400 при характерных значениях: $\Delta p_{\min}=27$ и $\Delta p_{\max}\hm=330$~МПа; $\Delta v_{\min}=0,033$ и $\Delta v_{\max}=0,435\ \text{см}^3/\text{г}$ показаны на рис.\,\ref{pic:ssi3d_d_ppi} (б).

\section{Заключение}

Сравнивая графики на рис.\,\ref{pic:ssi3d_d_ppi} (а-б) можно заметить, что на качественном уровне результаты перколяционного моделирования $d_{\Delta v}(p|\pi)$ демонстрируют удовлетворительное согласование с данными порометрических испытаний $\Delta p^*(\Delta v^*)$.
К примеру, вполне ясно, что кривая интрузии соответствует меньшему значению показателя Минковского, чем кривая экструзии: $d_{\Delta v}(p|\pi)_i<d_{\Delta v}(p|\pi)_e$, где $\pi_i<\pi_e$.
Для показанных на рис.\,\ref{pic:ssi3d_d_ppi} (б) кривых наиболее близкие значения показателей Минковского составят $\pi_i\approx \frac{1}{4}$ и $\pi_e\approx 1$.
С вероятностной точки зрения это означает, что средняя вероятность протекания для узлов из $(1, \pi)$-окрестности внутреннего узла трёхмерной перколяционной решётки при построении кластера экструзии $\Osr{p}_1=\frac{22p}{39}$ оказывается в $\frac{4752}{2219}\approx 2,14$ раза выше, чем при построении кластера интрузии $\Osr{p}_{\frac{1}{4}}=\frac{2219p}{8424}$, что и обеспечивает более высокие значения фрактальной размерности $d$ при тех же долях достижимых узлов $p$.
С физической точки зрения это приводит к выводу об увеличении средних значений эффективного гидравлического диаметра при переходе от интрузии к экструзии несмачивающей жидкости из поровых каналов \cite{moskaleff.bnak.2013.hysteresis}.


\end{document}